\documentclass[onecolumn,letterpaper,11pt,draftclsnofoot]{IEEEtran}
\usepackage[]{graphicx}

\usepackage{amsmath}
\usepackage{tabularx}
\usepackage{graphicx}
\usepackage{subfigure}
\usepackage{amssymb}
\usepackage{epstopdf}
\usepackage{color}
\usepackage{algorithm,algpseudocode}

\usepackage{cite}
\usepackage{url}

\begin{document} 
  
  \bstctlcite{IEEEexample:BSTcontrol}
  
  \title{\huge{A Games-in-Games Approach to Mosaic Command and Control Design of Dynamic Network-of-Networks for Secure and Resilient Multi-Domain Operations }}
\author{Juntao Chen and Quanyan Zhu\\ Department of Electrical and Computer Engineering, Tandon School of Engineering\\ New York University, Brooklyn, NY 11201 USA
\thanks{Further author information: (Send correspondence to Quanyan Zhu) Juntao Chen: E-mail: jc6412@nyu.edu; Quanyan Zhu: E-mail: qz494@nyu.edu, Telephone: 1-646-997-3371 }}

\maketitle

\begin{abstract}
This paper presents a games-in-games approach to provide design guidelines for mosaic command and control that enables the secure and resilient multi-domain operations. Under the mosaic design, pieces or agents in the network are equipped with flexible interoperability and the capability of self-adaptability, self-healing, and resiliency so that they can reconfigure their responses to achieve the global mission in spite of failures of nodes and links in the adversarial environment. The proposed games-in-games approach provides a system-of-systems science for mosaic distributed design of large-scale systems. Specifically, the  framework integrates three layers of design for each agent including strategic layer, tactical layer, and mission layer. Each layer in the established model corresponds to a game of a different scale that enables the integration of threat models and achieve self-mitigation and resilience capabilities. The solution concept of the developed multi-layer multi-scale mosaic design is characterized by Gestalt Nash equilibrium (GNE) which considers the interactions between agents across different layers. The developed approach is applicable to modern battlefield networks which are composed of heterogeneous assets that access highly diverse and dynamic information sources over multiple domains. By leveraging mosaic design principles, we can achieve the desired operational goals of deployed networks in a case study and ensure connectivity among entities for the exchange of information to accomplish the mission.
\end{abstract}


\section{INTRODUCTION}
\label{sec:intro}  

Security and resilience of networked systems become increasingly critical nowadays due to the prevailing adversarial threats from both cyber and physical domains \cite{chen2017dynamic}. A number of approaches have been proposed in literature to enhance the system performance under adversarial environment through strategic trust \cite{pawlick2017strategic,pawlick2019istrict,zhu2012guidex,fung2016facid}, resilient control \cite{chen2015resilient,chen2017interdependent,xu_game-theoretic_2017,xu2017secure,xu2015cyber,xu2015secure}, moving target defense \cite{jajodia2011moving,zhu2013game,maleki2016markov,jafarian2012openflow}, proactive deception \cite{al2019autonomous,jajodia2016cyber,Huang2019book,pawlick2018modeling,Pawlick2018Dissertation,pawlick2017game}, and contracts and insurances \cite{chen2016optimal,zhang2017bi,chen2017,chen2018linear}. With the adoption of Internet of things (IoT) devices and information and communications technologies (ICTs), different systems are integrated together, creating network-of-networks (NoN) \cite{kurian2017electric,zimmerman2017conceptual,zimmerman2018network}. On one hand, NoN improves the system dependability  and interoperablity \cite{Chen_ACC}. On the other hand, the network interdependency introduces new challenges for the system operator to maintain the NoN performance as the interconnection provides extra opportunity for the propagation of attacks from one network to another, e.g., through lateral movement in advanced persistent threat (APT) \cite{huang2019adaptive}.  Traditional defensive strategies for networked systems are no longer sufficient in this emerging NoN framework. Therefore, our goal in this paper is to propose an efficient and flexible way to achieve mission objectives while ensuring the security and resilience of NoN through a new paradigm called \textit{mosaic design}. The associated concept of mosaic warfare has been recently proposed by DARPA \cite{darpa}.

 Mosaic distributed system design refers to engineering agents with flexible interoperability and the capability of \textit{self-adaptability}, \textit{self-healing}, and \textit{resiliency}. Specifically, systems can achieve its objective when one node goes away or fails \cite{Chen_CDC,chen2019-games}. Furthermore,
systems can respond to other systems in a non-deterministic/stochastic way and increases the composability and modularity of the system design. For example, agents can randomly arrive and respond in a stochastic but structured way to other agents in an uncertain environment. However, the structured randomness leads to emerging system behaviors that manifest desirable properties for the objective of entire mission.
Systems that have such properties are easily composable and resilient-by-design. 
Without a pre-planned integration among agents, the agents can adapt their response and reconfigure their own systems based on the type of agents that they interact with. For example, in the decision-making of battlefield scenario, the unmanned ground vehicle (UGV) network should intelligently coordinate its actions with the heterogeneous unmanned aerial vehicle (UAV) network and the soldier network in a self-adaptive manner. Thus, in the paradigm of mosaic design, agents can be easily composed to achieve a prescribed objective through an unprescribed path. In addition,
under the adversarial environments, the agents can reconfigure their response and roles to achieve the global mission in spite of failures of nodes and communication links. Returning to the battlefield example, a single removal of UGV or UAV should not interrupt the action of other agents, and the whole system should still be operable when one piece is missing to achieve the global mission. These features of agile self-recovery ability and autonomous composability are the epicenter of the mosaic designs.

Mosaic design is a migration from a pre-defined protocol for distributed systems that aim to achieve a single objective. Classical design has a prescribed objective and then uses a top-down design methods to decentralize the operations. For example, the operator of the entire battlefield first designs optimal strategies for the agents globally and then inform each agent how to act based on their local information. However, when one agent leaves the battlefield which modifies the system, the previous designed strategy is not globally optimal anymore. Therefore, the loss of one piece will loose the entire effect in the classical top-down design. 
Mosaic design is also different from the classical deterministic bottom-up approach in which agents are programmed to behave in a designed way offline which losses the adaptivity.

 In this work, we develop a games-in-games approach to provide a system-of-systems science for mosaic distributed design of large-scale systems. Different from previous works in designing resilient operational strategy for interdependent networks based a single game \cite{huang2017factored,huang2017large,huang2018factored,huang2018distributed}, the games-in-games approach allows an automated composition of systems to achieve flexible interoperability \cite{zhu2015game,chen2019-games}. Agents can adapt to their neighboring ones and integrate themselves into the environment.
  The game-theoretic approach also enables the integration of threat models and achieve self-mitigation/resilience capabilities.

 \textit{Related Work}: Game-theoretic approaches have been extensively adopted for resilient control of networked system and critical infrastructures \cite{chen2016game,chen2017dynamic,chen2017interdependent,huang2017factored,huang2018distributed,chen2019dynamicgame}. To analyze strategic interactions between attackers and defenders, a large number of works have focused on the security modeling and design through game-theoretic frameworks \cite{huang2017large,zhang2017bi,huang2019adaptive,huang2018factored,zhu2015game,chen2018security}. Furthermore, researchers have also used game-theoretic methods to enable decentralized multi-layer network/network-of-networks design \cite{Chen_ACC,Chen_CDC,chen2017heterogeneous,chen2019}. Due to the integration between heterogeneous system components, interdependent security and trust mechanisms become critical and they have been addressed through game-theoretic methods from a system-of-systems perspective \cite{chen2016optimal,chen2017,pawlick2017strategic,pawlick2019istrict}. When the number of agents in the network grows, secure and resilient control needs to incorporate the feature of large-scale complex systems, e.g., multi-layer IoT networks and epidemic networks \cite{chen2019-epidemics,Farooq,farooq2018multi,hayel2017epidemic}.

\section{Games-in-Games Approach for Mosaic Design}
In this section, we develop a games-in-games framework which enables multi-layer and multi-scale decision-making over network systems. Then, we design the mosaic control based on this games-in-games framework.

\subsection{Games-in-Games Framework}

The games-in-games principle is a framework that provides a theoretical underpinning and a guideline for mosaic control designs. 
Specifically, the proposed games-in-games approach integrates three layers of design for each agent: strategic layer, tactical layer, and mission layer. 
At the strategic layer, the agents learn and respond to their environment quickly to unanticipated events such as attacks, disruptions, and changes of other agents.
At the tactical layer, the agents plan for a longer period of time by taking into account the long-term interactions with the environment and other agents. The agents can make a goal-oriented planning at each stage.
At the mission layer, the agents develop a stage-by-stage planning of multi-stage objectives to achieve the mission despite the uncertainties and online changes.

Each layer in the established model corresponds to a game of a different scale. 1): At the strategic layer, a game associated with an agent describes its interaction with an adversary, e.g. a jammer, a spoofer, or a sudden loss of neighboring node. Solutions to this game can prepare nodes for unanticipated attacks and secure the agents. 2): At the tactical layer, an $N$-person dynamic game describes the longer-term interactions among cooperative agents, each seeking control policies to achieve individual stage objectives. The individual control would lead to achieving global objectives such as connectivity and network formation.
3): At the mission layer, each agent plans at each stage their stage objectives at the tactical layer. This planning is obviously under a lot of uncertainties and need to be achieved in a moving-horizon way.

\begin{figure}[t]
  \centering
    \includegraphics[width=.92\columnwidth]{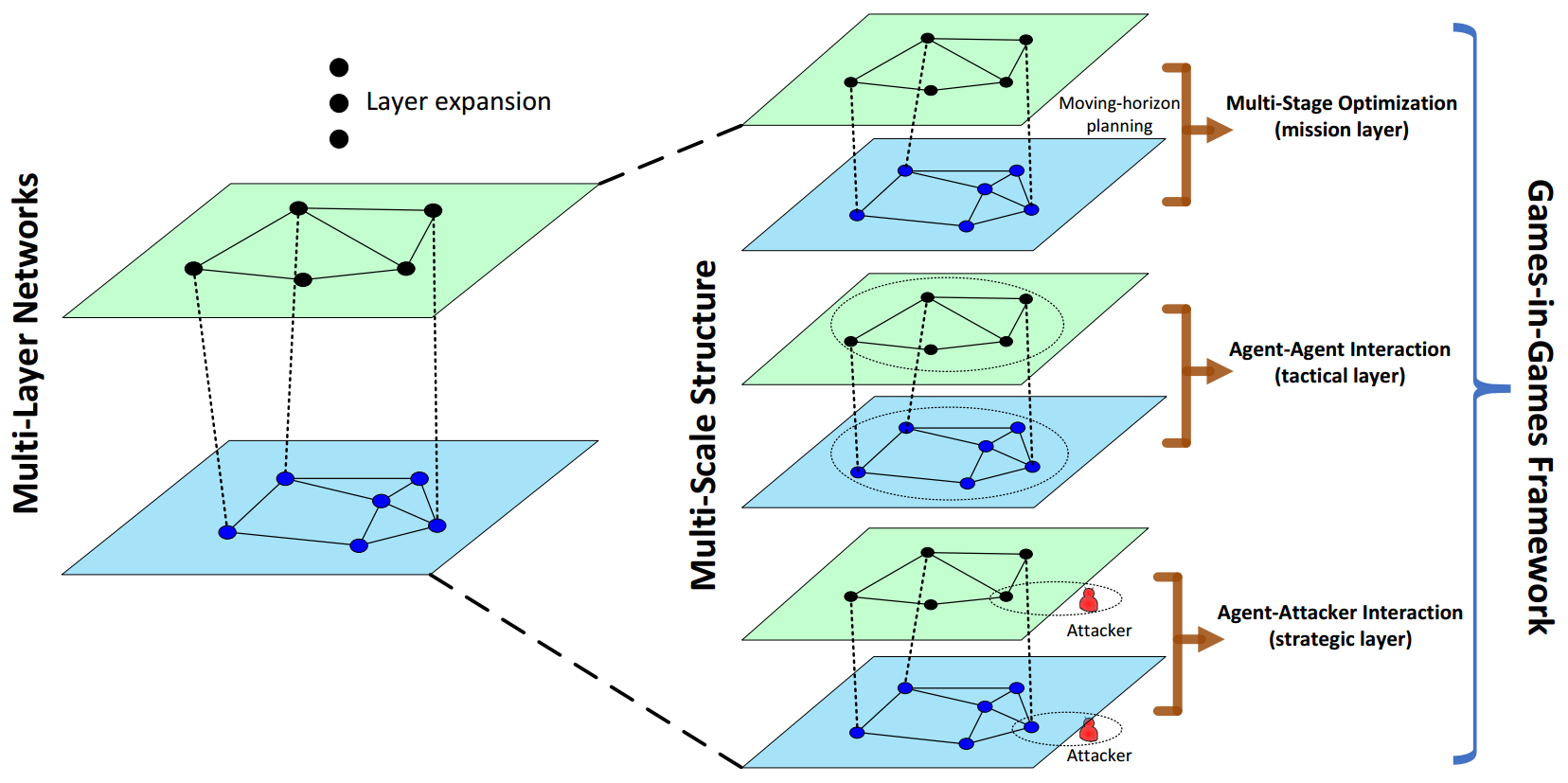}
      \caption[]{Games-in-Games framework for mosaic command and control design of secure and resilient networks-of-networks. The games-in-games framework contains three layers: strategic layer for attack and disruption consideration of each agent; tactical layer for interaction consideration between agents within and across different layers at each stage; and mission layer for moving-horizon planning to achieve multi-stage objective. Games at different layers can be composed together, leading to a flexible mosaic control design. }
  \label{mosaic_framework}
\end{figure}

In sum, games-in-games framework describes a multi-layer and multi-scale game-theoretic framework.
Furthermore, the games at each layer can be composed together. For example, an $N$-person game can be composed with an $M$-person game to create an $N+M$-person game.  Such composition leads to a resolution of the games at each layer.
The games across the layers can also be composed together. For example, an $N$-person tactical layer game is nested in an $N$-person mission-layer game. In addition, the security game at the strategic layer can be nested in the tactical layer games. For clarity, the games-in-games framework is illustrated in Fig. \ref{mosaic_framework}.

\subsection{Mosaic Command and Control Design}
The developed games-in-games framework can be adopted to address the mosaic control design as the composability of the framework provides agility required by the mosaic control objective. This framework is inherently secure and resilient by design.  First, the games-in-games framework anticipates the attack behavior and designs a control policy that would prepare to defend against the anticipated attacks. The framework provides a clean-slate design and provides a built-in security for each system component that would lead to security of the integrated system. Second,
the games-in-games framework enables each system to respond to the unanticipated events at each time instant. Each agent can respond to events that inflict damages on the agent and go through a self-healing process that can recover itself from the attacks and failures if possible. If the full recovery is not achievable, the agents will develop control strategies that will allow a graceful performance degradation.
Therefore, the multi-layer mosaic control enables the agents to achieve mission despite of failures, uncertainties, and unstructured behaviors.
Note that the mosaic control is a fully integrated design which differentiates itself from current existing designs in which only partial aspects are considered, e.g., security, but not all key issues.

The solution concept of the developed multi-layer and multi-scale mosaic design is characterized by Gestalt Nash equilibrium (GNE) \cite{chen2018security,chen2019-TIFS}.
Nash equilibrium provides a solution concept to a well defined static or dynamic game in strategic or extensive forms. We extend this solution concept to GNE for games-in-games framework where multiple games can be composed to capture heterogeneous interactions among different types of players. The GNE solution concept follows the definition of NE and describes an equilibrium concept in which no agent has incentives to deviate away from not only the local game, which captures the local agent-agent level interactions,  but also the composed game, which captures the global system-system level interactions.
The development of GNE provides a solution concept for multi-scale interactions, provides a way to assess system-level performance, and enables the design of mosaic control systems.

Mosaic control design is suitable for multi-domain operations (MDO) \cite{perkins2017multiA,perkins2017multiB}. MDO refers to a cross-domain integration of information and assets across air, space, sea, land, and cyber domains to provide a holistic situational awareness and decision-making.  
We can leverage mosaic control design to provide a framework to develop a modular, functional, and composable design of command and control systems that can autonomously achieve the mission objectives.

 \begin{figure}[t]
   \centering
     \subfigure[]{
     \includegraphics[width=0.4\columnwidth]{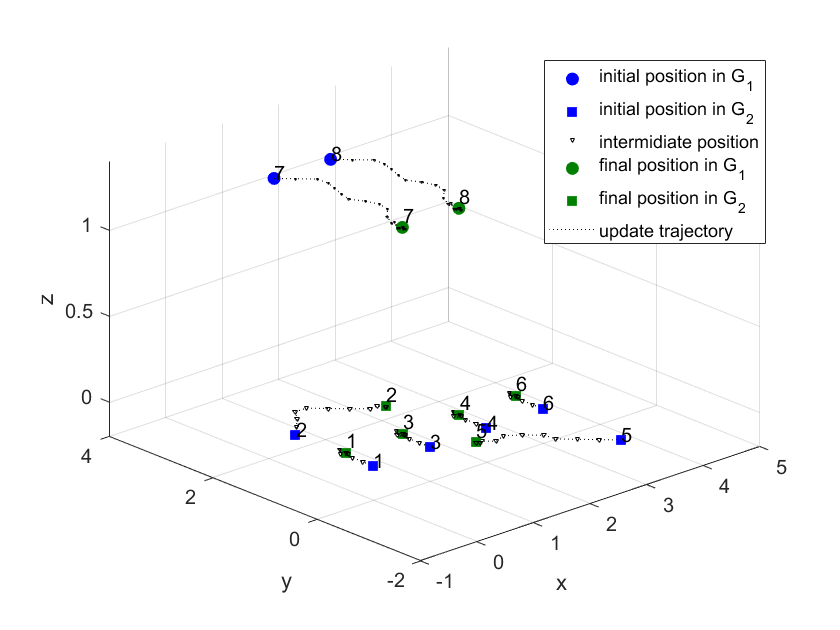}} 
       \subfigure[]{
     \includegraphics[width=0.4\columnwidth]{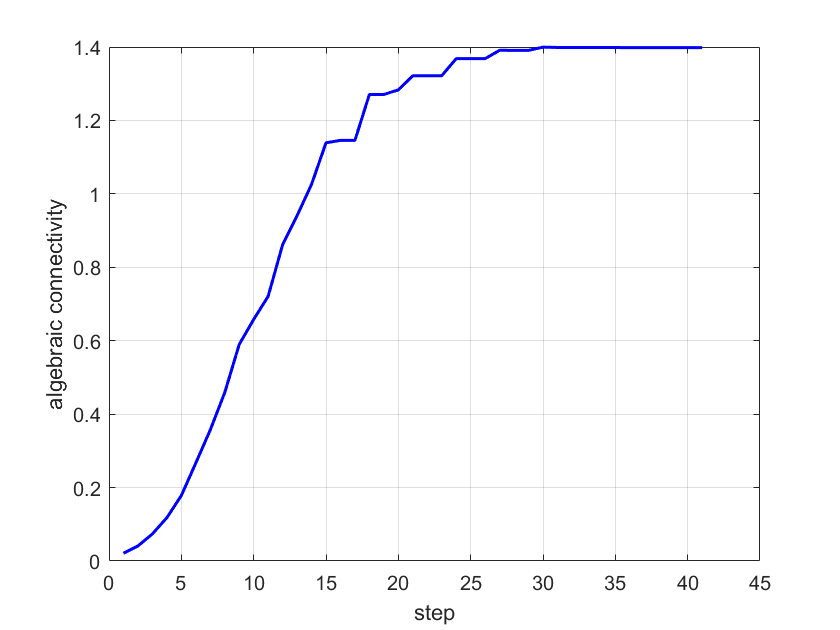}} 
   \caption[]{(a) shows the evolutionary configuration of secure MAS network at each step with the consideration of jamming attack. (b) shows the corresponding network connectivity.}
   \label{base_case}
 \end{figure}

\subsection{Examples and Results}
\begin{figure}[t]
  \centering
    \subfigure[]{
    \includegraphics[width=0.35\columnwidth]{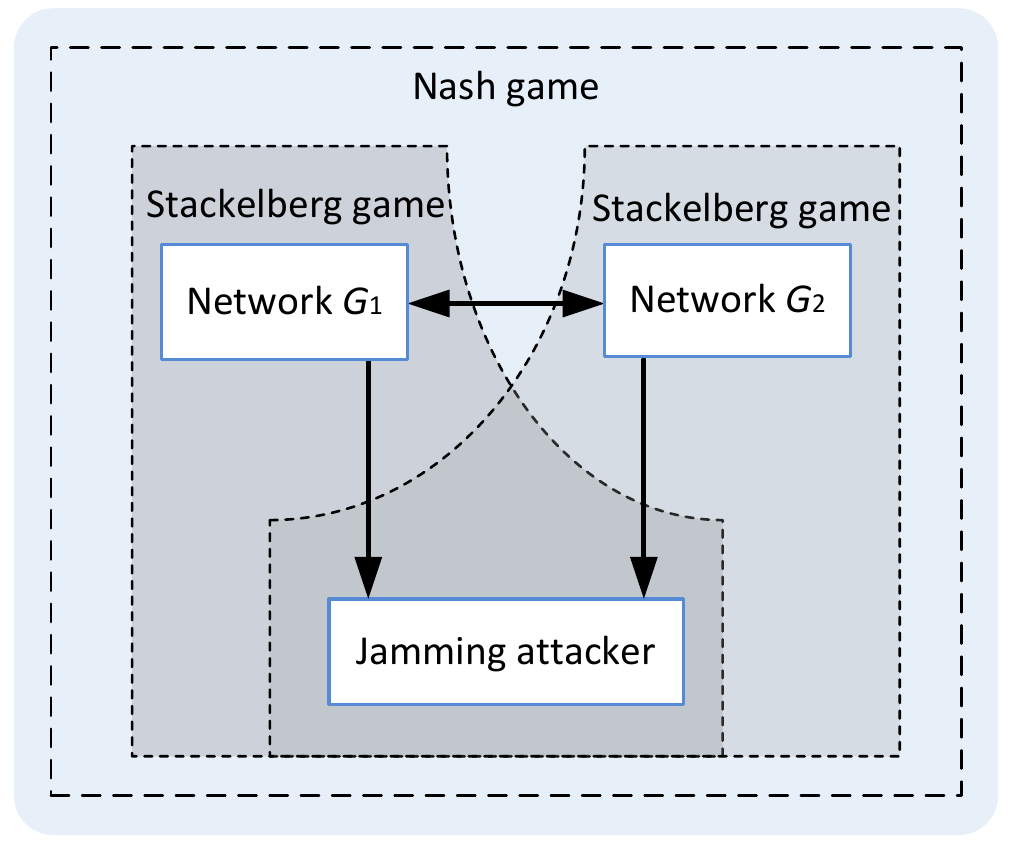}}
  \subfigure[]{
    \includegraphics[width=0.45\columnwidth]{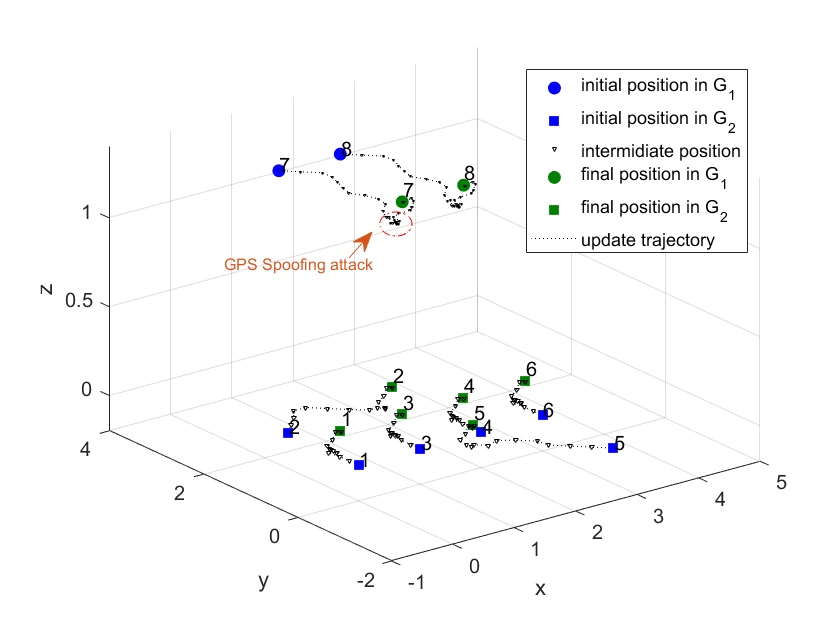}} 
      \subfigure[]{
    \includegraphics[width=0.45\columnwidth]{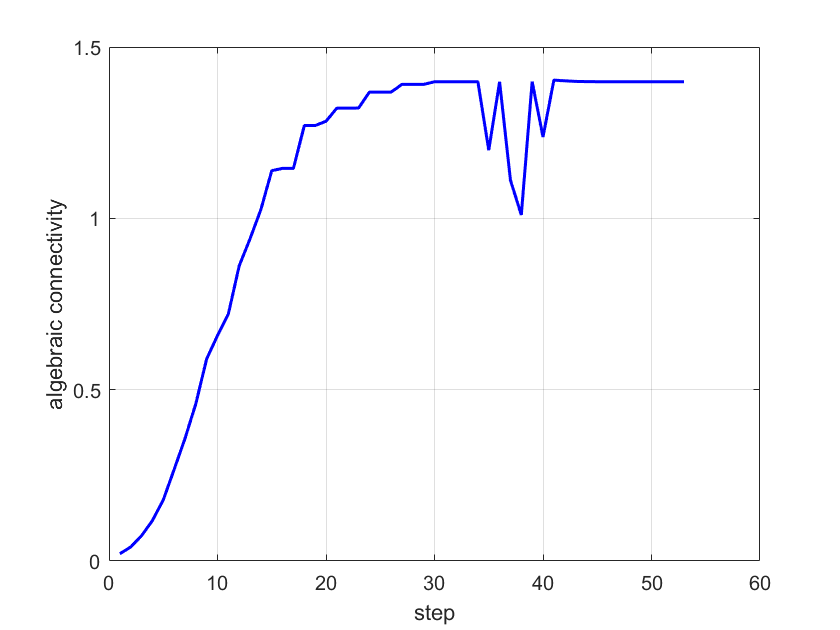}\label{spoofing_connectivity_equilibrium_attack}}
  \caption[]{(a) depicts a games-in-games framework for two-layer autonomous systems. (b) shows the iterative configuration of a two-layer autonomous network under mosaic control. (c) shows the corresponding network connectivity. The spoofing attack launches at step 35 and lasts for 6 steps. The network recovers and reaches a GNE quickly afterward.}
  \label{spoofing_case_equilibrium_attack}
\end{figure}

In this subsection, we study a case study of a multi-layer network of autonomous systems \cite{Chen_CDC}, in which UAVs and UGVs act collaboratively, intelligently, and adaptively to achieve a high connectivity. Furthermore, the designed decentralized MDO command and control algorithms enable a synchronized response for each layer to respond to others to maintain real-time connectivity despite the adversarial environment. Here, we present numerical results of a two-layer mobile autonomous systems using mosaic design principles. Maintaining connectivity between different agents is critical which improves the network situational awareness \cite{chen2017heterogeneous,chen2019}. In the case studies, the objective of two network operators is to optimize the global network algebraic connectivity by anticipating the existence of adversary \cite{Chen_CDC}.
As shown in Fig. \ref{base_case}, the network is robust to jamming attack and maintains connectivity with the presence of a jammer at every step which demonstrates the security of the mosaic control algorithm. In addition, as depicted in \ref{spoofing_case_equilibrium_attack}, the nodes can respond quickly to the spoofing attack and achieve agile resilience through the proposed control design. Interested readers can find more results and discussions of case studies in \cite{chen2019-games,chenbook}.

\bibliographystyle{IEEEtran}
\bibliography{IEEEabrv,references}

\end{document}